\documentstyle[12pt]{JHEP}

\title{Unbroken versus broken mirror world: a tale of two vacua}

\author{R. Foot, H. Lew and R. R. Volkas \\ School of Physics 
\\ Research Centre for High
Energy Physics \\
The University of Melbourne \\ Victoria 3010 Australia \\
\email{foot@physics.unimelb.edu.au} \\
\email{Henry.Lew@dsto.defence.gov.au}\\
\email{r.volkas@physics.unimelb.edu.au}}

\abstract{If the Lagrangian of nature respects parity 
invariance then there are two
distinct possibilities: either parity is unbroken by the vacuum or it is
spontaneously broken. We examine the two simplest phenomenologically consistent
gauge models which have unbroken and spontaneously
broken parity symmetries, respectively. 
These two models have a Lagrangian of the
same form, but a different parameter range is chosen in
the Higgs potential. They both predict the
existence of dark matter and can explain the MACHO events. However, the models
predict quite different neutrino physics. Although both have light mirror
(effectively sterile) neutrinos, the ordinary-mirror neutrino mixing 
angles are unobservably
tiny in the broken parity case. The minimal broken parity model therefore cannot
simultaneously explain the solar, atmospheric and LSND data. By contrast, the
unbroken parity version can explain all of the neutrino anomalies. 
Furthermore, we
argue that the unbroken case provides the most 
natural explanation of the neutrino
physics anomalies (irrespective of whether evidence from the LSND 
experiment is
included) because of its characteristic maximal mixing prediction.}

\begin{document}

\section{Introduction}

Parity is a natural candidate for a symmetry in particle physics. The fact that
experiments have established the $V-A$ nature of the weak interactions does not mean
that parity cannot be a symmetry of the vacuum. This is due to there being no unique
definition of the parity transformation in quantum field theory. Parity, by
definition, takes $\vec{r}$ to $-\vec{r}$ and it also transforms left-handed fermion
fields to right-handed fermions fields. However, since there are many left- and
right-handed fermion fields, there is no unique definition. It is certainly true,
though, that the standard model has no definition of parity which can be a symmetry of
its Lagrangian. Thus, if parity is a symmetry of the Lagrangian of nature, new
physics must exist. 

It was realised some time ago that the simplest phenomenologically consistent gauge
model with both a parity symmetric Lagrangian and vacuum could be constructed by
postulating the existence of a mirror sector \cite{flv} (see also Ref.\cite{gl}).
The idea was discussed earlier, before the advent of gauge theories, in
Ref.\cite{ly} and some cosmological implications were considered in
Refs.\cite{blin,turn,cg}.

Suppose that for each known particle there exists a corresponding mirror particle.
The mirror particles interact with other mirror particles in the same way that the
ordinary particles interact with the other ordinary particles, except that mirror
weak interactions are right-handed ($V+A$) instead of left-handed ($V-A$). The
doubling of the particle spectrum which parity invariance suggests is not
particularly radical in the historical context of particle physics. In particular
the repetitive generation structure, the existence of antiparticles and the colour
tripling of quarks are all familiar examples.  Of these examples, Dirac's
prediction of antiparticles from the requirement of Lorentz invariance is perhaps
most analogous to the prediction of mirror particles from parity invariance (parity
being an improper Lorentz transformation). Although the number of particles is
doubled, it turns out that the number of parameters increases by only two (in the
minimal version), making it the simplest (non-trivial) known extension to the
standard model.

It turns out that in the minimal model, which has one Higgs doublet and one mirror
Higgs doublet, the Higgs potential allows two non-trivial minima \cite{fl}. One
minimum leaves parity unbroken by the vacuum, and the other has it spontaneously
broken. The purpose of this paper is to compare and contrast the physics resulting
from the two different vacuum solutions (which effectively can be considered two
different theories). Both theories are phenomenologically consistent and are
therefore possible extensions to the standard model. 
The mirror atoms/baryons in both cases
provide plausible candidates for the dark matter in the universe. However, neutrino
physics should be able to distinguish the two theories as we will show.

\section{The Lagrangian and the two vacua}

Let us begin by briefly reviewing the model (for full details see Ref.\cite{flv}).
Consider the minimal standard model Lagrangian ${\cal L}_1$. This Lagrangian is not
invariant under the usual parity transformation so it seems parity is violated.
However, this Lagrangian may not be complete. If we add to ${\cal L}_1$ a new
Lagrangian ${\cal L}_2$ which is just like ${\cal L}_1$ except that all left-handed
(right-handed) fermions are replaced by new right-handed (left-handed) fermions
which feel new interactions of the same form and strength, then the theory
described by ${\cal L} = {\cal L}_1 + {\cal L}_2$ is invariant under a parity
symmetry which induces ${\cal L}_1 \leftrightarrow {\cal L}_2$. In addition to
these Lagrangian terms, there may also be parity invariant terms which couple or
mix ordinary matter with mirror matter. We label this part of the Lagrangian as
${\cal L}_{int}$. The terms in ${\cal L}_{int}$ are very important since they lead
to non-gravitational interactions between ordinary and mirror matter and hence
allow the idea to be experimentally tested in the laboratory. These terms also
contain the only new parameters of the model. In the minimal case, only two terms
exist in ${\cal L}_{int}$ because of the constraints of gauge invariance, parity
invariance and renormalisablility. For this reason the minimal model has only two
new parameters beyond those of the minimal standard model.

The gauge symmetry of the theory 
is\footnote{Of course, grand unified alternatives such as
$SU(5)\otimes SU(5)$, $SO(10)\otimes SO(10)$ and so on  
are also possible. Such alternatives predict
similar low energy physics (see for example Refs.\cite{gl,cf}).
The idea may also be compatible with string theory ($E_8 \otimes
E_8$) and with large extra dimensions \cite{zur}.}
\begin{equation}
SU(3) \otimes SU(2) \otimes U(1) \otimes
SU(3)' \otimes SU(2)' \otimes U(1)'.
\label{gsym}
\end{equation}
There are two sets of fermions, the ordinary particles
(denoted without primes in the following)
and their mirror images -- the mirror particles
(denoted with primes). The fields
transform under the gauge group of Eq.(\ref{gsym}) as
\begin{equation}
\begin{array}{ll}
f_L \sim (1, 2, -1)(1, 1, 0),\quad & f'_R \sim (1, 1, 0)(1, 2, -1),\\
e_R \sim (1, 1, -2)(1, 1, 0),\quad & e'_L \sim (1, 1, 0)(1, 1, -2),\\
q_L \sim (3, 2, 1/3)(1, 1, 0),\quad & q'_R \sim (1, 1, 0)(3, 2, 1/3),\\
u_R \sim (3, 1, 4/3)(1, 1, 0),\quad & u'_L \sim (1, 1, 0)(3, 1, 4/3),\\
d_R \sim (3, 1, -2/3)(1, 1, 0),\quad & d'_L \sim (1, 1, 0)(3, 1, -2/3),
\end{array}
\end{equation}
with generation indices suppressed. The Lagrangian is invariant
under the discrete $Z_2$ parity symmetry 
defined by\footnote{By virtue of standard $CPT$ invariance the theory also possesses
time reversal invariance (which can be defined to be the
product of the above parity transformation with the usual
$CPT$ transformation) \cite{flv}. The theory is therefore invariant under the full 
Poincar\'e group, including all improper Lorentz transformations.}
\begin{equation}
\begin{array}{c}
\vec{r} \rightarrow -\vec{r},\quad  t \rightarrow t,\\
G^{\mu} \leftrightarrow G'_{\mu},\quad
W^{\mu} \leftrightarrow W'_{\mu},\quad B^{\mu}
\leftrightarrow B'_{\mu},\\
f_L \leftrightarrow \gamma_0 f'_R,\quad e_R \leftrightarrow
\gamma_0 e'_L,\quad q_L \leftrightarrow \gamma_0 q'_R,\quad
u_R \leftrightarrow \gamma_0 u'_L,\quad d_R \leftrightarrow
\gamma_0 d'_L, \end{array}
\end{equation}
where $G^{\mu}$ $(G'^{\mu})$, $W^{\mu}$ $(W'^{\mu})$
and $B^{\mu}$ $(B'^{\mu})$ are the gauge bosons of
the $SU(3)$ $[SU(3)']$,
$SU(2)$ $[SU(2)']$ and  $U(1)$ $[U(1)']$
gauge forces respectively.
The minimal model contains two Higgs doublets,
\begin{equation}
\phi \sim (1, 2, 1)(1, 1, 0),\quad  \phi' \sim (1, 1, 0)(1, 2, 1),
\end{equation}
which are also parity partners.

The most general renormalisable
Higgs potential can be written in the form
\begin{equation}
V(\phi, \phi') = \lambda_+ (\phi^{\dagger} \phi
+ \phi'^{\dagger} \phi' - 2u^2)^2 + \lambda_- (
\phi^{\dagger} \phi - \phi'^{\dagger} \phi')^2,
\label{pot1}
\end{equation}
where $\lambda_{\pm}$ and $u^2$ are real parameters. 
This Higgs potential furnishes only two vacua when $u^2 > 0$.
For the region
of parameter space defined by $\lambda_{\pm} > 0$ the vacuum
expectation values
(VEVs) of $\phi$ and $\phi'$
are equal and so parity is unbroken. (With the Higgs potential
written in the above form this is manifest because
the potential is non-negative, and equal to zero
for this vacuum solution).
Gauge invariance allows us to write the vacuum in the 
form
\begin{equation}
\langle \phi \rangle = \langle \phi' \rangle =
\left( \begin{array}{c}
0\\
u
\end{array}\right).
\label{ttt}
\end{equation}

Interestingly, a simple analysis shows that there is a qualitatively different minimum
which occurs in the region of parameter space defined by
\begin{equation}
\lambda_+ + \lambda_- > 0\quad {\rm and}\quad   \lambda_- < 0.
\end{equation}
The vacuum solution now has
one VEV nonzero and the other zero \cite{fl}.
In this region of parameter space, parity is broken spontaneously. 
This can be made manifest
by rewriting the Higgs potential in terms
of the parameters $\lambda_1 \equiv 
\lambda_+ + \lambda_-$ and  $\lambda_2 \equiv
- 4\lambda_-$,
\begin{equation}
V(\phi, \phi') = \lambda_1 (\phi^{\dagger} \phi +
\phi'^{\dagger} \phi' - u_B^2)^2 +  \lambda_2 (\phi^{\dagger}\phi
\phi'^{\dagger}\phi')
\label{pot2}
\end{equation}
where
\begin{equation}
u_B^2 \equiv \frac{2\lambda_+ u^2}{\lambda_+ + \lambda_-}.
\end{equation}
Written in this way, for $\lambda_{1,2} > 0,$ the vacuum solution
\begin{equation}
\langle \phi \rangle = u_B, \quad \langle \phi' \rangle = 0,
\label{vac1} 
\end{equation}
or
\begin{equation}
\langle \phi \rangle = 0, \quad \langle \phi' \rangle = u_B,
\label{vac2}
\end{equation}
is manifest since the Higgs potential is non-negative,
and equal to zero
for these two degenerate vacua. In the spontaneously broken 
parity model we adopt the solution of Eq.(\ref{vac1})
for definiteness. Notice that the parameters must be adjusted to make $u$ and $u_B$
both numerically equal to the observed electroweak symmetry breaking VEV (but not
simultaneously, of course). 

Let us examine the mirror particle spectrum
of these two theories.
In the unbroken mirror model 
[the one with the vacuum of Eq.(\ref{ttt})]
all of the mirror fermions and mirror gauge bosons
have the same mass as the
corresponding ordinary fermions and gauge bosons.

In the broken mirror model
[the one with the vacuum of Eq.(\ref{vac1})] it seems that
the mirror fermions and the mirror gauge bosons are all massless
because they all couple to
$\phi'$ which has a vanishing VEV.
However, dynamical effects from mirror
QCD induced condensation will induce a small mass for
the mirror weak bosons, $W'$ and $Z'$, as well as a
tiny VEV for $\phi'$, and hence
tiny masses for the mirror fermions. The latter effect is induced by the Yukawa
coupling terms $h_{q} \overline{q}'_R q'_L \phi' + H.c.$, where $h_q$ is the
Yukawa coupling constant for quark flavour $q$. Mirror QCD interactions
cause
the mirror quark bilinears to have nonzero vacuum expectation values: $\langle
\overline{q}' q' \rangle = \Lambda'^3$ where $\Lambda' \sim 100$ MeV is the mirror
QCD
dynamical chiral symmetry breaking scale \cite{fl}. The resulting linear term in
$\phi'$
effectively contributes to the Higgs
potential so as to induce a small VEV, given by $\langle \phi' \rangle \simeq
h_t \Lambda'^3/m^2_{\phi'}$. Note that the largest contribution to the linear term
comes from mirror top quark condensation due to the relatively large Yukawa
coupling constant $h_t$.
Further details
have been given in Ref.\cite{fl}.

The result is the following particle spectrum:
\begin{itemize}
\item The $W'$ and $Z'$ bosons will have masses of order
$\Lambda' \sim 100$ MeV.
\item The four physical mirror
scalars (two complex scalars $\phi'^{+}$ and $\phi'^0$) will be approximately
degenerate with mass
$\sqrt{\lambda_2}\ u_B$.
\item The mirror fermions $\psi'$ will have masses given by
\[
m_{\psi'} = k m_\psi \qquad \hbox{where} \qquad
k \equiv {\langle \phi' \rangle \over \langle \phi \rangle }
\sim {g^2\over 2}{m_t \Lambda'^3\over m_W^2 m_{\phi'}^2}
\]
where $g$ is the $SU(2)$ gauge coupling constant,
$m_W$ is the ordinary $W$ boson
mass, and $m_\psi$ is the mass of the corresponding ordinary 
fermion $\psi$. Using
the rough estimates 
$g^2 \sim 10^{-1}$,
$m_t \sim m_W  \sim 10^2$ GeV and
$\Lambda' \sim 0.1$ GeV, we see that
$k \sim 10^{-6}(\hbox{GeV}/m_{\phi'})^2$.
For example, if $m_{\phi'} \sim 10^2$ GeV,
then $k \sim 10^{-10}$.
This means that the masses of the mirror fermions can range from
about $10$ eV for the mirror top-quark to $10^{-4}$ eV for the
mirror electron. As this example illustrates, we expect $k$ to be
quite small ($\stackrel{<}{\sim} 10^{-10}$). 
Its precise value, however, cannot be specified because $m_{\phi'}$
depends on the free parameter $\lambda_2$.
\end{itemize}

\section{Phenomenology}

If the solar system is dominated by the ordinary particles
(and this is expected theoretically as we will discuss shortly),
then both of these theories agree with present experiments. (An
extension to
incorporate nonzero neutrino masses 
and mixings must be performed to get agreement
with the neutrino deficit experiments.) These ideas can be tested in the laboratory
because of the terms in ${\cal L}_{int}$.
In the simplest case that we are considering at the moment
(where ${\cal L}_1$ is the minimal standard model Lagrangian), there are just
two gauge and parity invariant
and renormalisable terms in ${\cal L}_{int}$. They are: the Higgs potential
term
\begin{equation}
2(\lambda_+ - \lambda_-) \phi^{\dagger} \phi
\phi'^{\dagger} \phi'
\label{higgsmixing} 
\end{equation}
contained within Eq.(\ref{pot1}), and
the gauge boson kinetic mixing term 
\begin{equation}
\omega F_{\mu \nu} F'^{\mu \nu}
\label{kineticmixing}
\end{equation}
where $F_{\mu \nu} = \partial_{\mu} B_{\nu} -
\partial_{\nu} B_{\mu}$ is the field strength tensor for the
$U(1)$ gauge boson, and $F'_{\mu\nu}$ is its mirror analogue.

The main phenomenological effect of the term in Eq.(\ref{higgsmixing}) is to modify
the interactions of the Higgs boson. This effect will be tested
if or when a Higgs boson is discovered. 
For the unbroken case, the implications of the
Higgs mixing term have been discussed in Refs.\cite{flv,flv2,ray}.
In the spontaneously broken case, the ordinary physical neutral Higgs boson $H$
couples via cubic and quartic terms to the mirror Higgs doublet $\phi'$: 
\begin{equation}
{\cal L}_{H\phi'} = 
2\sqrt{2}(\lambda_1 + \frac{1}{2} \lambda_2) u_B H \phi'^{\dagger}
\phi' + 
(\lambda_1 + \frac{1}{2} \lambda_2) H^2 \phi'^{\dagger} \phi'.
\label{brokenHiggsints}
\end{equation}
An experimental lower bound can be derived for $\lambda_1$ by observing that
\begin{equation}
\lambda_1 = {m_{H} \over 2 u_B} \stackrel{>}{\sim} 0.2
\label{lambdaprimebound}
\end{equation}
where the experimental lower bound on the standard model
Higgs boson mass of roughly $90$ GeV has been used. This bound 
is valid even if
$H$ decays invisibly \cite{bound} into mirror matter (via
$H \to \phi'^{\dagger} \phi'$).
Thus the cubic and quartic ordinary -- mirror Higgs 
interactions are necessarily quite large in the 
spontaneously broken model since $\lambda_2$, being a positive quantity, cannot
be fine-tuned to make $\lambda_1 + (1/2)\lambda_2 \simeq 0$. 
It is also interesting to observe that
there is no ordinary -- mirror Higgs mass mixing in the spontaneously
broken model. These remarks will be important when we discuss cosmological
implications below.

The main phenomenological effect of the kinetic mixing term in
Eq.(\ref{kineticmixing}) is to give small electric charges to the mirror partners
of the ordinary charged fermions \cite{flv,gl,bob}. In the case where the parity
symmetry is unbroken, photon -- mirror photon kinetic mixing leads to
orthopositronium -- mirror orthopositronium oscillations \cite{gl}. There is
interesting evidence that these oscillations have actually been observed
experimentally \cite{gn}. However, this tantalising result needs experimental
confirmation.
 
If neutrinos are massive,
then this will allow an important window on the
mirror world which was first realised 
in 1991 \cite{flv2}. 
This is because ${\cal L}_{int}$ can contain neutrino
mass terms which mix the ordinary and mirror 
matter.\footnote{Note that if electric and mirror electric charges are conserved,
then it is not possible for ${\cal L}_{int}$ to contain
mass terms mixing the charged fermions of ordinary matter
with those of mirror matter.}
A remarkable, but very simple, result is that if ordinary and mirror
neutrinos mix at all then the resulting neutrino 
oscillations must be maximal
if the parity symmetry is unbroken \cite{flv2,fv}.
To see this consider
the electron neutrino. If there is no mirror matter,
and if intergenerational mixing is small as in the quark sector, then
the weak eigenstate electron neutrino is approximately
a single mass eigenstate. However
if mirror matter exists, then there will be a mirror
electron neutrino $\nu'_e$.
The most general mass matrix consistent
with parity conservation [Eq.(5)] is
\begin{equation}
{\cal L}_{mass} = [\bar \nu_e, \bar \nu'_e]
\left(\begin{array}{cc}
m&m'\\
m'&m
\end{array}\right)\left[
\begin{array}{c}
\nu_{e}\\
\nu'_e
\end{array}\right] + H.c. \end{equation}
where the masses can be taken as real without loss of generality.
This mass matrix has the above form whether the neutrinos
are Majorana or Dirac states.
Diagonalising, we easily obtain that
the weak eigenstates $\nu_e$ and $\nu'_e$ are each maximally mixed
combinations of mass eigenstates:
\begin{equation}
\nu_{e} = {(\nu_1^+ + \nu_1^{-}) \over \sqrt 2},\quad
\nu'_{e} = {(\nu_1^+ - \nu_1^{-}) \over \sqrt 2}, 
\end{equation}
where $\nu_1^+$ and $\nu_1^{-}$ are the mass eigenstates.
Thus, the effect of ordinary matter mixing with
mirror matter is very dramatic. No matter how small
(or large)
the mass interaction term is, the mixing is maximal.
This result immediately leads to a beautiful and natural
explanation of the neutrino physics anomalies \cite{flv2,fv}:
The solar anomaly is due to maximal $\nu_e \to \nu'_e$ oscillations, the
atmospheric anomaly is due to maximal $\nu_\mu \to \nu'_{\mu}$
oscillations, while the LSND anomaly can be accomodated through small
intergenerational
mixing between the first and second families.
Maximal mixing can be considered a ``smoking gun'' for
unbroken parity invariance in nature!

For the broken parity case, the maximal mixing feature between
the ordinary and mirror neutrinos is completely lost. In fact, the mixing angles
between any ordinary neutrino and its mirror partner will be suppressed by the tiny
parameter $k \equiv \langle \phi' \rangle/\langle \phi \rangle$. To see this, it is
simplest
to think in terms of effective operators. At the dimension-5 level the relevant
terms are 
\begin{equation}
{\cal L}_{eff} = \frac{a}{M} \left[ \overline{f}_L \phi^c\  
\phi^{\dagger} (f_L)^c + \overline{f'}_R \phi'^c\
\phi'^{\dagger} (f'_R)^c \right]
+ \frac{b}{M} \overline{f}_L \phi^c\
(\phi'^c)^{\dagger} f'_R
 + H.c.,
\end{equation}
where $M$ is a high scale, perhaps the seesaw scale.
The resulting effective low
energy Majorana mass matrix yields 
a mixing angle of order $k$.\footnote{One of the
authors (RRV) thanks K S Babu for
emphasising this point in conversation.}

So, although the minimal broken mirror model provides a nice rationale for light
effectively sterile neutrinos, then can play no observable role in neutrino
phenomenology! In particular, the solar, atmospheric and LSND anomalies cannot be
simultaneously resolved. Furthermore, the possible scenarios for explaining the
solar and atmospheric results involve the ordinary neutrinos only, and thus have no
novel features. The predictivity of the unbroken mirror model for neutrino physics
is totally absent in the broken version.

\section{Astrophysics and cosmology}

In both theories, the lightest mirror baryon is stable
and provides a natural candidate for the
inferred dark matter of the universe. 
Mirror matter will behave like cold dark
matter on large scales, but the self interactions will
modify the physics on small scales.
In the unbroken case, the interactions and masses of the mirror 
particles are completely analogous to the ordinary particles.
The lightest mirror baryon in this case is, of course,
the mirror proton.
This case has been studied in Refs.\cite{blin,turn,hod} where it
was argued that the mirror particles provide
a plausible candidate for the dark matter of the universe.
It was also shown that the mirror matter may be
segregated from ordinary matter on 
distance scales much larger than the solar system but may
be well mixed on the scale of galaxies. 
Interestingly, there is observational evidence that some of the dark matter
in our galaxy is in the form of invisible stars.
This evidence comes from the MACHO candidates obtained
from the microlensing observations of stars in the Large Magellanic Cloud 
\cite{alcock}.
These observations support the existence of mirror
matter since mirror matter naturally forms star mass sized
objects \cite{ll}.
There are also a number of other interesting cosmological
implications of the unbroken mirror world which are discussed in Ref.\cite{other}.

In the spontaneously broken mirror model, we expect 
the lightest mirror baryon to
be (mirror) electrically neutral. We will call it the mirror neutron. 
It is the
lightest mirror baryon because the mirror up quark -- down 
quark mass difference is
negligible ($\ll$ eV), so the mirror proton -- mirror neutron mass 
difference is
dominated by the electromagnetic corrections. This means that the mirror 
proton would be expected to be a few MeV {\it heavier} than 
the mirror neutron. Thus,
mirror neutrons would be stable and form the mirror dark matter. 
In the broken model, the mirror pions will have very
small mass $\sim \sqrt{k}m_{\pi} \stackrel{<}{\sim} 1$ keV.
This will significantly increase the range of the mirror
strong interactions. In view of this it seems quite plausible
that the interactions of the mirror neutrons of the spontaneously broken
mirror model will be strong enough and dissapative
enough to cause the collapse
of a mirror neutron gas into
into mirror stars/planets and the like. The spontaneously
broken model may therefore also explain the mysterious MACHO
population inferred from the microlensing observations (although
perhaps not quite as naturally as the unbroken model).

Of course, observations suggest that the 
amount of cold dark matter needed
may be significantly greater than the
amount of ordinary baryons in the universe.
This feature is not inconsistent with an exact
mirror symmetry when one realises that
microscopic symmetries do not always lead
to macroscopic symmetries. Thus, the number of mirror baryons
in the universe need not be exactly the same as the number
of ordinary baryons even if the microscopic parity symmetry 
is unbroken.
It is, however, most plausible for the number of
mirror baryons in the universe to be of the same order of
magnitude as the number of ordinary baryons.
On the other hand, the moderately successful Big Bang Nucleosynthesis (BBN) 
predictions suggest that the temperature of the mirror
plasma was somewhat lower (a factor of two would
suffice) than the temperature of the ordinary plasma during the BBN epoch.
The temperature difference between the ordinary and mirror
matter can be ascribed to unknown physics at early times
or divine intervention. Specific mechanisms based on inflation have been
studied in Refs.\cite{turn,hod,berez}. 

Of course for consistency it is important
that the non-gravitational interactions between the ordinary and mirror
particles be small enough so that they do not
populate the
mirror sector and wipe out the temperature difference.
We will consider in turn the three generic effects of ${\cal L}_{int}$:
photon -- mirror photon kinetic mixing, neutrino -- mirror neutrino 
mass mixing, and Higgs boson
-- mirror Higgs boson interactions and mixing. 

A stringent constraint from BBN was derived some years ago on the
photon - mirror photon kinetic mixing parameter $\omega$ \cite{cg}.
Little more needs to be said about this here, except to note that the bound
is only as robust as the BBN scenario itself.

The BBN bounds on neutrino to mirror neutrino oscillations in the symmetric mirror
model have been discussed at length in the literature \cite{fvlong}. The important
point is that large relic neutrino asymmetries will be generated by ordinary to
mirror oscillations (just as for ordinary to sterile oscillations) \cite{ftv}
provided that the oscillation parameters are in the appropriate range. The large
asymmetries can then suppress mirror neutrino production because the matter effect
depresses the effective mixing angle \cite{fvprl}.
In the broken mirror model, mirror neutrino
production is automatically very suppressed by the tiny mixing angles whose origin 
was explained above.

We now need to derive consequences for BBN due to Higgs boson -- mirror
Higgs boson
coupling and mixing. The consequences actually depend on cosmology at temperature
scales
much higher than those of BBN. It is simplest to discuss this issue within the
context of inflationary cosmology, even though the existence of an inflationary
epoch has yet to be observationally established. Suppose first of all that the
reheating temperature $T_{RH}$ of the ordinary plasma was at least several hundred
GeV. Recall that inflationary scenarios can be constructed to provide asymmetric
reheating for the ordinary and mirror plasmas: $T'_{RH} < T_{RH}$
\cite{turn,hod,berez}.
Will subsequent
ordinary -- mirror Higgs interactions force the mirror plasma temperature $T'$ to
equal the ordinary plasma temperature $T$? In this case,
electroweak symmetry is presumably restored and the ordinary Higgs doublet $\phi$
is a component of the ordinary plasma. The relevant process to consider is
therefore
$\phi^* \phi \to \phi'^* \phi'$ scattering. If we want the rate for
this scattering process to be less than the expansion rate of the universe for all
temperatures above the phase transition temperature, the stringent
constraint \cite{cf, hen}, 
\begin{equation}
\lambda_+ - \lambda_- \equiv \lambda_1 + \frac{1}{2} \lambda_2 \stackrel{<}{\sim}
3\times 10^{-6} \sqrt{\frac{m_\phi}{TeV}},
\label{higgsbound1}
\end{equation}
must be satisfied. For the symmetric mirror model, this constraint 
can be met by
requiring the two positive parameters $\lambda_+$ and $\lambda_-$ to almost
cancel.\footnote{
This is actually technically natural because the combination
$\lambda_+ - \lambda_-$ is just the coupling between the ordinary and mirror
scalars
which if zero would increase the symmetry of the theory (separate
global Lorentz groups for the ordinary and mirror sectors).}
Once this has been done, the bound of Eq.(\ref{higgsbound1}) also suffices
to prevent ordinary -- mirror plasma thermal equilibration by Higgs boson processes
occurring after electroweak 
symmetry breakdown. For the broken mirror model, this
constraint can never be satisfied, because $\lambda_2$ is positive and $\lambda_1$ 
is bounded from below as per Eq.(\ref{lambdaprimebound}). Thus, if
$T_{RH}$ is higher than the Higgs boson mass scale, then the ordinary and
mirror plasmas must come to thermal equilibrium, making the model incompatible with
standard BBN.

The relevance of the bound in Eq.(\ref{higgsbound1}) was recently re-examined in
Ref.\cite{ray} for the symmetric mirror
model case. It was pointed out that the $T_{RH}$ could be much lower than the
electroweak phase transition temperature. If so, then the bound in
Eq.(\ref{higgsbound1}) is irrelevant. Instead, a weaker bound pertains which arises
from Higgs boson -- mirror Higgs boson mass mixing after symmetry breaking. See
Ref.\cite{ray} for full details, and in particular for a discussion of the dramatic
Higgs boson phenomenology that could be discovered by the Large Hadron Collider if
this scenario is correct.

Using similar reasoning, the broken mirror case can be made compatible with BBN if
the reheating temperature is less than the Higgs boson mass scale. However, after
the electroweak phase transition\footnote{Note that this phase transition can be
eliminated by having a sufficiently low $T_{RH}$.} the cubic term in
Eq.(\ref{brokenHiggsints}) will induce the process 
\begin{equation}
b \overline{b} \to t' \overline{t}' t' \overline{t}'.
\end{equation} 
Using
dimensional analysis, the rate for this process is
\begin{equation}
\Gamma = a (\lambda_1 + \frac{1}{2} \lambda_2)^2 h_t^4 \frac{m_b^2}{m^4_{H}
m^8_{\phi'}} T^{11}
\end{equation}
where $a$ is a numerical factor expected to be in the range $0.1 - 10$. Requiring
this rate to be less than the expansion rate ${\cal H} \sim T^2/M_P$, where $M_P$
is the
Planck mass, we obtain the rough bound
\begin{equation}
\lambda_1 + \frac{1}{2} \lambda_2 \stackrel{<}{\sim} \frac{m^2_{H}
m^4_{\phi'}}{h_t^2 m_b \sqrt{M_P}} T_{RH}^{-\frac{9}{2}} \sim 100 \left(
\frac{m_{\phi}}{100\
\hbox{GeV}} \right)^6 \left( \frac{T_{RH}}{\hbox{GeV}} \right)^{-\frac{9}{2}},
\label{Higgsbound2}
\end{equation}
where $m_{\phi}$ is now a generic Higgs or mirror Higgs boson mass scale. We see
that
the bound is quite weak for reasonable $m_{\phi}$ with the reheating
temperature in the $10$'s of GeV range.

As our final comment on the cosmology of the broken mirror model, we address the
issue of a possible domain wall problem. Superficially, it appears that this model,
like other models with spontaneously broken discrete symmetries, would present a
serious domain wall problem if the electroweak phase transition actually took
place.
While this potential problem can be solved by postulating a sufficiently low
reheating temperature as considered above, we will argue that for this particular
model the domain walls are necessarily unstable even if the electroweak phase
transition occurred, 
provided that the radiation-dominated phase of the universe was never hotter than
$m_{\phi}$.
This is because of our expectation
that the ordinary and mirror plasmas will have different temperatures in order to
achieve successful BBN. The asymmetric temperatures will cause asymmetric finite
temperature corrections to the Higgs potential. Although the two parity breaking
vacua are degenerate at zero temperature, they will not be degenerate for any
finite $T$ and $T'$ if $T \neq T'$. The different energy densities in the domains
on either side of a wall should cause the wall to collapse due to the ensuing
pressure differential. Of course, because of the bound in Eq.(\ref{higgsbound1}),
the reheating temperature must still be constrained to be less than $m_{\phi}$,
otherwise $T' < T$ cannot hold at lower temperatures.

\section{Next-to-minimal broken mirror model}

Several years after the unbroken and spontaneously broken mirror models were
introduced, another type of mirror model was discussed in Ref.\cite{bm}
(and some cosmological implications discussed in 
Ref.\cite{nm}). These
authors modified the minimal models \cite{flv,fl} discussed here by postulating
an additional scalar particle. This allows $k \equiv \langle \phi'
\rangle/\langle \phi \rangle$ to be
a free parameter \cite{bm}, in contrast to both the unbroken and broken
incarnations of the minimal model. While this model is distinct from the minimal
mirror models, it is not quite as predictive because it is has more free parameters
(especially the mirror symmetry breaking scale). 
It has also been used to address
the neutrino problems, (with the $\nu_e \to \nu_s$ small angle MSW 
solution\cite{bar} to the solar neutrino problem and $\nu_\mu \to \nu_\tau$
solution to the atmospheric neutrino anomaly)\cite{bm}, 
as well as the
dark matter and MACHO problems \cite{ll}.

\section{Concluding remarks}

The hypothesis of mirror matter allows one to write down a phenomenologically
tenable gauge model with parity (and in fact the full Poincar\'e group) 
as an exact symmetry of both the Lagrangian and the
vacuum state of nature \cite{flv}. Interestingly, the same Lagrangian supplies also
a parity asymmetric vacuum state for a certain range of parameters \cite{fl}. The
phenomenology, astrophysics and cosmology of both incarnations of the minimal
mirror model have been examined in this paper.

One of the most urgent tests of 
these ideas lies with neutrino physics. The unbroken or
symmetric mirror model features maximally mixed pairs of ordinary and mirror
neutrinos \cite{flv2,fv}. For terrestrial experiments, mirror neutrinos are
effectively sterile states. The most natural solution to the atmospheric neutrino
anomaly uses maximal $\nu_\mu \to \nu'_\mu$ oscillations, while the solar neutrino
anomaly is most attractively solved through maximal $\nu_e \to \nu'_e$ oscillations
\cite{flv2,fv}. From an experimental perspective, this is a combined maximal
$\nu_\mu \to \nu_s$ and $\nu_e \to \nu'_s$ scenario. While neither of these
solutions currently provide a perfect fit to the experimental data, they are
broadly consistent with the atmospheric \cite{yas} and solar experiments
\cite{crock}. Observe that if it turns out that the atmospheric neutrino anomaly is
due to $\nu_\mu \to \nu_\tau$ oscillations then this, by itself, cannot rule
out the unbroken mirror matter model. It just means that the
$\nu_\mu \to \nu'_\mu$ oscillation length -- which is controlled by a free
$\Delta m^2$ parameter-- is much greater than $10,000$ km for
atmospheric neutrino energies relevant to the experiments. 

The forthcoming neutral current measurement from SNO will be a crucial test of the
$\nu_e \to \nu'_e$ solution to the solar neutrino problem. Efforts to discriminate
between the $\nu_{\mu} \to \nu_s$ and $\nu_{\mu} \to \nu_{\tau}$ solutions to the
atmospheric neutrino anomaly, both from the MINOS and CERN -- Gran Sasso long
baseline experiments and possibly from SuperKamiokande itself, will also be
important. While SuperKamiokande claims to disfavour the $\nu_{\mu} \to \nu_s$ or
$\nu'_{\mu}$ solution on the basis of a preliminary analysis \cite{vigans}, we have
to wait for a
detailed account of their analysis procedure in order to judge the robustness of
this preliminary conclusion. The Miniboone experiment will also be important,
because its ability to check the LSND result \cite{lsnd} will also confirm or
disconfirm the
latter's indirect evidence for a light sterile neutrino. If the participation of a
light sterile state in
the neutrino anomalies is confirmed by future data, then the minimal {\it broken}
mirror model will certainly be ruled out.

\acknowledgments
This work was supported by the Australian Research Council. RRV acknowledges useful
discussions with K. S. Babu that took place during the 2nd Tropical Workshop on
Particle Physics and Cosmology in San Juan, Puerto Rico.


\begin{thebibliography}{99}

\bibitem{flv}
R. Foot, H. Lew and R. R. Volkas, \plb{272}{1991}{67}.

\bibitem{gl}
S. L. Glashow, \plb{167}{1986}{35}.

\bibitem{ly}
T. D. Lee and C. N. Yang, \pr{104}{1956}{256};
I. Kobzarev, L. Okun and I. Pomeranchuk, \sjnp{3}{1966}{837}; 
M. Pavsic, {\it Int. J. Theor. Phys.} {\bf 9} (1974) 229.

\bibitem{blin}
S. I. Blinnikov and M. Yu. Khlopov, \sjnp{36}{1982}{472};
{\it Sov. Astron.} {\bf 27} (1983) 371. 

\bibitem{turn}
E. W. Kolb, M. Seckel and M. S. Turner, {\it Nature} {\bf 514} (1985) 415.

\bibitem{cg}
E. D. Carlson and S. L. Glashow, \plb{193}{1987}{168}.

\bibitem{fl} R. Foot and H. Lew, \hepph{9411390} (July, 1994).

\bibitem{cf}
M. Collie and R. Foot, \plb{432}{1998}{134}.

\bibitem{zur}
Z. Silagadze \mpla{14}{1999}{2321}; G. C. Joshi, Private communication;
See also N. Arkani-Hamed et al., \hepph{9911386}.

\bibitem{flv2} R. Foot, H. Lew and R. R. Volkas, \mpla{7}{1992}{2567}.

\bibitem{ray}
A. Yu. Ignatiev and R. R. Volkas, \hepph{0005238}.

\bibitem{bound}
L3 collaboration, M. Acciarri et al., \hepex{0004006}.

\bibitem{bob}
B. Holdom, \plb{166}{1986}{196}.

\bibitem{gn}
R. Foot and S. N. Gninenko, \plb{480}{2000}{171};
see also S. N. Gninenko, \plb{326}{1994}{317}.

\bibitem{fv}
R. Foot and R. R. Volkas, \prd{52}{1995}{6595};
R. Foot, \mpla{9}{1994}{169}.

\bibitem{hod}
H. M. Hodges, \prd{47}{1993}{56}. 

\bibitem{alcock}
C. Alcock et al., (MACHO Collab), \astroph{0001272};
T. Lasserre (EROS Collab), \astroph{9909505};
L. V. E. Koopmans and A. G. de Bruyn, \astroph{0004112}.

\bibitem{ll}
Z. Silagadze, {\it Phys. At. Nucl.} {\bf 60} (1997) 272; 
S. Blinnikov, \astroph{9801015}; 
R. Foot, \plb{452}{1999}{83}; 
R. Mohapatra and V. Teplitz, \plb{462}{1999}{302}.

\bibitem{other}
M. Yu. Khlopov et al., {\it Sov. Astron.} {\bf 35} (1991) 21;
S. Blinnikov, \astroph{9902305}; \astroph{9911138};
Z. Silagadze, \hepph{9908208}; \hepph{0002255};
G. Matsas et al., \hepph{9810456};
N. F. Bell and R. R. Volkas, \prd{59}{1999}{107301};
R. Foot, \plb{471}{1999}{191}.
R. R. Volkas and Y. Y. Y. Wong, {\it Astropart. Phys.} {\bf 13} (2000) 21;
N. F. Bell, \plb{479}{2000}{257};
R. M. Crocker, F. Melia and R. R. Volkas, \astroph{9911292}.

\bibitem{berez}
V. Berezinsky and A. Vilenkin, \hepph{9908257}.

\bibitem{fvlong}
R. Foot and R. R. Volkas, \prd{61}{2000}{043507}; 
{\it Astropart. Phys.}{\bf 7} (1997) 283.

\bibitem{ftv}
R. Foot, M. J. Thomson and R. R. Volkas, \prd{53}{1996}{5349};
R. Foot and R. R. Volkas, \prd{55}{1997}{5147}; \prd{56}{1997}{6653};
P. Di Bari, P. Lipari and M. Lusignoli, \hepph{9907548} (to appear
in Int. J. Mod. Phys. A).

\bibitem{fvprl}
R. Foot and R. R. Volkas, \prl{75}{1995}{4350}.

\bibitem{hen}
H. Lew, unpublished.

\bibitem{bm}
Z. Berezhiani and R. N. Mohapatra, \prd{52}{1995}{6607}.

\bibitem{nm}
Z. G. Berezhiani, A. Dolgov and R. N. Mohapatra, \plb{375}{1996}{26};
Z. G. Berezhiani, {\it Acta Phys. Polon.} {\bf B27} (1996) 1503.

\bibitem{bar}
V. Barger et al. \prd{43}{1991}{1759}.

\bibitem{yas}
R. Foot, R. R. Volkas and O. Yasuda, \prd{58}{1998}{013006}; 
\prd{57}{1998}{1345};
P. Lipari and M. Lusignoli, \prd{58}{1998}{073005}; 
N. Fornengo, M. C. Gonzalez-Garcia and J. W. F. Valle, \hepph{0002147}.

\bibitem{crock}
R. M. Crocker, R. Foot and R. R. Volkas, \plb{465}{1999}{203}; 
R. Foot, \hepph{0003189} (to appear in Phys.Lett.B);
R. Foot and R. R. Volkas, \hepph{9510312}.

\bibitem{vigans}
M. Vigans, talk at 2nd Tropical Workshop on Particle Physics and Cosmology, San
Juan, Puerto Rico, May 1-6 2000.

\bibitem{lsnd}
LSND Collaboration, C. Athanassopoulos et al., \prl{75}{1995}{2650};
\prl{77}{1996}{3082}; \prl{82}{1999}{2430}.

\end{thebibliography}
\end{document}